\documentclass[prd,twocolumn,showpacs,preprintnumbers,floats,floatfix,apsrev,amsmath,amsfonts,nofootinbib]{revtex4-1}
\usepackage{color}
\usepackage{graphicx}
\usepackage{dcolumn}
\usepackage{txfonts}
\usepackage{multirow}
\usepackage{subcaption}
\usepackage{natbib}
\usepackage{hyperref}
\usepackage{hypernat}
\usepackage{verbatim}

\newcommand{\dd}{{\rm d}}
\newcommand{\be}{\begin{equation}}
\newcommand{\en}{\end{equation}}
\newcommand{\bea}{\begin{eqnarray}}
\newcommand{\ena}{\end{eqnarray}}
\newcommand{\G}{\mathrm{G}}

\begin{document}

\title{Spacetime Singularities in Generalized Brans-Dicke Theories}
\author{G.~Brando}\email{gbrando@cbpf.br}
\affiliation{CBPF - Centro Brasileiro de
Pesquisas F\'{\i}sicas, Xavier Sigaud st. 150,
zip 22290-180, Rio de Janeiro, Brazil.}

\author{F.~T.~Falciano}\email{ftovar@cbpf.br}
\affiliation{CBPF - Centro Brasileiro de
Pesquisas F\'{\i}sicas, Xavier Sigaud st. 150,
zip 22290-180, Rio de Janeiro, Brazil.}

\author{L.~F.~Guimar\~aes}\email{lfog@cbpf.br}
\affiliation{CBPF - Centro Brasileiro de
Pesquisas F\'{\i}sicas, Xavier Sigaud st. 150,
zip 22290-180, Rio de Janeiro, Brazil.}

\date{\today}

\begin{abstract}

We study the formation of classical singularities in Generalized Brans-Dicke theories that are natural extensions to Brans-Dicke where the kinetic term is modified by a new coupling function $\omega(\varphi)$. We discuss the asymptotic limit $\omega(\varphi)\rightarrow\infty$ and show that the system generically does not approach General Relativity. Given the arbitrariness of $\omega(\varphi)$, one can search for coupling functions chosen specifically to avoid classical singularities. However, we prove that this is not the case. Homogeneous and spherically symmetric collapsing objects form singularities for arbitrary coupling functions. On the other hand, expanding cosmological scenarios are completely free of Big Rip type singularities. In an expanding universe, the scalar field behaves at most as stiff matter, which makes these cosmological solutions asymptotically approach General Relativity.

\end{abstract}


\maketitle

\section{Introduction}

In a recent letter, Alexander et al.~\cite{alexander2016} investigate possible Higgs-like mechanisms to account for the emergence of gravity in extensions to General Relativity (GR). The main idea is that symmetry restoration (breaking) could ``turn off'' (``turn on'') gravity. The authors argue that a possible realization of this scenario is to identify the scalar field in Brans-Dicke (BD) theory with the Higgs-like field that undergoes symmetry restoration. At low curvature configurations the system must conform to GR but at high curvature regime a phase transition can drastically modify the dynamics. In particular, one might wonder if this type of Higgs-like mechanisms could dynamically avoid classical singularities.

BD theory is an attempt to incorporate Mach's Principle in a relativistic gravitational theory. This pioneer work by C. Brans and R. H. Dicke \cite{brans1961} can be seen as a scalar-tensor gravity theory where the extra scalar field is non-minimally coupled to the Ricci scalar. This coupling introduces an effective gravitational strength inversely proportional to the scalar field, $G_{eff} \sim \varphi^{-1}$. Recently, BD received much attention due to its phenomenological applications in cosmology~\cite{hrycyna2014,hrycyna2013,papagiannopoulos2016,alonso2016,roy2017}. The kinetic term associated with $\varphi$ carries a constant dimensionless parameter $\omega_{BD}$. In order for BD to be consistent with solar system astronomical experiments, the BD parameter must be large ($\omega_{BD} > 500$ see~\cite{will1993}). Effectively, this constrains BD to be observationally indistinguishable from GR in the weak field limit.

There are black hole solutions in BD~\cite{bronnikov1973,bronnikov1999,scheel1995,campanelli1993}, even though not with exactly the same features as in GR, and an analogue version of Birkhoff's theorem that proves that a static, spherically symmetric, asymptotically flat, vacuum solution in the Jordan frame is uniquely characterized by the Schwarzschild solution~\cite{faraoni2018}. Thus, it is certain that BD has singular solutions.

A straightforward generalization of BD is to promote $\omega$ to be a function of the scalar field as it is the case of generalized Brans-Dicke theories (GBD). In fact, BD can be seen as a particular example of GBD where the coupling function is fixed to be a constant. GBD was first introduced by Nordvedt \cite{nordtvedt1970}, where he analyze the post-Newtonian corrections to weak-field regime. Almost simultaneously, Wagoner~\cite{wagoner1970} also studied the weak-field limit considering the predictions for classical tests as light propagation and perihelion shift effects and the observable differences between the scalar and tensor components in gravitational waves.

Our goal in the present work is to analyze if GBD can dynamically avoid the formation of singularities. For that we shall discuss collapsing and cosmological solutions. In the next section we briefly describe GBD theories and the conditions for approaching the GR regime. In section \ref{MCGS} we discuss the necessary matching conditions and the geometrical setting for our dynamic system. In section \ref{CCO} and \ref{COSMO} we analyze, respectively, the collapsing and cosmological scenarios and in \ref{Conclusions} we conclude with some final remarks.

\section{Generalized Brans-Dicke Theories}\label{GBDT}

The GBD theory describes a family of generalized scalar-tensor theories. The action reads
\begin{equation}\label{genaction}
S=\int \dd x^4 \sqrt{-g}\left[\frac1{2\kappa}\left(\varphi R-\frac{\omega(\varphi)}{\varphi}\nabla_{\alpha}\varphi\nabla^{\alpha}\varphi
\right)+\mathcal{L}_{\mathrm{mat}}\right] \ .
\end{equation}
where $\kappa \equiv (8\pi\G)^{-1}$. The scalar field has been normalized such that for constant $\varphi=1$ one recovers GR. For each coupling function $\omega(\varphi)$ one has a different modified gravity theory.

The extra scalar degree of freedom motivates the use of this framework to describe inflationary models. In 1989, La and Steinhardt presented a new inflationary scenario, called extended inflation \cite{la1989, yoshimura1991,guth1992,kolb1990,holman1991}, based on the original Brans-Dicke theory. Using scalar-tensor theories of gravity with canonical kinetic term, Acetta and Steinhardt~\cite{steinhardt1990} investigated inflationary models for an arbitrary function of $\varphi$ coupled to the Ricci scalar. Additionally, Barrow and Maeda~\cite{barrow1990} used GBD theories (\ref{genaction}) to model an inflationary epoch. In contrast to pure BD framework, these last two approaches were named hyperextended inflation \cite{laylock1994,copeland1994,wands1995,mimoso1995,roberts1995}. Still in the context of cosmology, Barrow \cite{barrow1993i,barrow1993ii} used the action (\ref{genaction}) to study FLRW cosmological models. As already pointed out, the widespread use of this modified theory of gravity is due to its rich phenomenology, as seen by the arbitrary character of $\omega(\varphi)$, and the possible role that the scalar field may play in these scenarios.

From the action \eqref{genaction}, we can obtain the dynamic equations by varying it with respect to the metric and the scalar field:
\begin{align}
& G_{\mu \nu} = \frac{8 \pi}{\varphi} T^{(m)}_{\mu \nu} + \frac{\omega(\varphi)}{\varphi^{2}} \left( \nabla_{\mu} \varphi \nabla_{\nu} \varphi - \frac{1}{2} g_{\mu \nu} \nabla^{\alpha} \varphi \nabla_{\alpha} \varphi \right) \nonumber \\
&\quad \qquad +\frac{1}{\varphi} \left( \nabla_{\mu} \nabla_{\nu} \varphi - g_{\mu \nu} \Box \varphi  \right), \label{FulleqBD} \\
&\Box \varphi = - \frac{\varphi}{2 \omega} R - \frac{1}{2} \left( \nabla^{\alpha} \varphi \nabla_{\alpha} \varphi \right)  \left( \frac{1}{\omega} \frac{\dd \omega }{\dd \varphi} - \frac{1}{\varphi} \right), \label{eqsemtraco}
\end{align}
with $G_{\mu \nu} = R_{\mu \nu} - \frac{1}{2}g_{\mu \nu} R$ being the Einstein tensor. Taking the trace of equation (\ref{FulleqBD}), we can rewrite equation (\ref{eqsemtraco}) as:
\begin{equation}\label{FulleqKG}
\Box \varphi = \frac{1}{2\omega +3} \left( 8 \pi T^{(m)} - \frac{\dd \omega}{\dd \varphi} \nabla^{\alpha} \varphi \nabla_{\alpha} \varphi \right),
\end{equation}
where $T^{(m)}$ is the trace of the matter energy-momentum tensor.

By construction, the action is diffeomorphism invariant. Since all variables are dynamic fields, one immediately has conservation of energy-momentum. One can also directly check that by taking the divergence of \eqref{FulleqBD} and using \eqref{eqsemtraco}
\begin{equation}\label{consvT}
\nabla_\mu T^{(m)\, }{}^{\mu \nu}=0 \ . 
\end{equation}

In our analysis, the matter content is described by conventional barotropic fluids with constant equation of state
\begin{equation}\label{eqstate}
p=(\gamma-1)\rho \quad, \qquad 1\leq \gamma\leq 4/3\ .
\end{equation}

We shall also mention the relation between the Jordan and Einstein frames. As described above, GBD is written in the Jordan frame. As it is well known, a conformal transformation in the metric \cite{faroni2004,faraoni2011} take us to the Einstein frame where the action resembles the Einstein-Hilbert action but with the scalar field now non-minimally coupled to the matter fields. Some authors claim that GBD (including the original BD) is conformally equivalent to Einstein's theory. However, there is an intense debate in the literature concerning the physical equivalence of these two frameworks \cite{saltas2011,kamenschik2016,faraoni1999i,faraoni2007,faraoni1999ii,bhattacharya2017}. 

The collapse of a scalar field within the scope of GR has already been analyzed in the literature.
For example, the seminal work of Christodoulou~\cite{christodoulou1994} showed that a naked singularity can occur in the spherical collapse of a scalar field, while Choptuik~\cite{choptuik1992} developed a numerical study for a massless scalar field considering a family of solutions with the property that a critical parameter, $p^\ast$, indicates the formation of black holes. Since then several works studied the critical behavior and the properties of the gravitational collapse of scalar fields~\cite{wang1997i,wang1997ii, banerjee2017, chakrabarti2017,zhang2015,torres2014,ganguly2011, bhattacharya2011}.

In the present work we will concentrate only on the avoidance or formation of singularities within GBD for the gravitational collapse and cosmological scenarios. Furthermore, in our case the scalar field is part of the geometry and not a matter field, as considered in the works mentioned above. The difference on the causal structure and the critical behavior for a collapsing system in GBD is a very interesting subject that we shall analyze in the future.

\subsection{General Relativity Limit}\label{GBDT:GRLim}
The experimental success of GR compels any alternative theory of gravity to look for a regime that is indistinguishable from GR. BD theory is an example where solutions can be continuously deformed into GR solutions as one makes the parameter $\omega_{BD} \rightarrow \infty$. In particular, it is argued \cite{will1993, reasenberg1973} that Solar System time-delay experiments sets a lower bound of $\omega_{BD} > 500$. A typical argument to support that BD theory approaches GR in the $\omega_{BD}\rightarrow\infty$ goes as follows \cite{weinberg1972}. The BD dynamics reads
\begin{align}
	\square \varphi=&\frac{T^\mu{}_\mu}{2\omega_{BD}+3}\label{BDeq1}\\
	G_{\mu\nu}=&\frac{8\pi T_{\mu\nu}}{\varphi}
	+\frac{1}{\varphi}\left(\nabla_\mu \nabla_\nu \varphi-g_{\mu\nu}\square\varphi \right)
	\nonumber\\
	&
	+\frac{\omega_{BD}}{\varphi^2}\left(\nabla_\mu\varphi \nabla_\nu \varphi-\frac12g_{\mu\nu}\nabla^\alpha\varphi \nabla_\alpha\varphi \right)\label{BDeq2}\ .
\end{align}
When the parameter $\omega_{BD} \gg 1$, equation \eqref{BDeq1} seems to show that $\square \varphi=\mathcal{O}\left(\omega_{BD}^{-1}\right)$, hence 
\begin{align}
	\varphi=&\frac1G +\mathcal{O}\left(\frac{1}{\omega_{BD}}\right)
	\label{BDeq12}\\
	G_{\mu\nu}=&\kappa T_{\mu\nu}+\mathcal{O}\left(\frac{1}{\omega_{BD}}\right)\label{BDeq22}
\end{align}
where $G$ is Newton's gravitational constant. Equation \eqref{BDeq22} goes to Einstein's equations in the limit $\omega_{BD}\rightarrow\infty$. However, it has been shown \cite{anchordoqui1998,nariai1968,banerjee1985,banerjee1986,banerjee1997,hanlon1972,matsuda1972,romero1993a,romero1998,romero1993b,paiva1993a,paiva1993b,scheel1995} that several exact solutions of BD theory do not tend to the corresponding GR solution in the limit $\omega_{BD}\rightarrow\infty$. The asymptotic behavior differs from equation \eqref{BDeq12} decaying as 
\begin{equation}\label{BDeq13}
\varphi=\frac1G +\mathcal{O}\left(\frac{1}{\sqrt{\omega_{BD}}}\right)\quad .
\end{equation}

In this case the last term of equation \eqref{BDeq2} does not go to zero, which breaks the GR limit. This feature is closely related with the conformal invariance of the matter content. Conformal invariance of BD theory, when considering pure gravitational systems or with a traceless matter content, implies that the action functional is invariant under an one-parameter abelian group of transformation that changes the value of the parameter $\omega_{BD}$. Faraoni has shown \cite{faraoni1999iii} that the set of all BD theories connected by this symmetry operation forms an equivalence class. Then, conformal invariant rescale the BD parameter such that the limit $\omega_{BD}\rightarrow\infty$ can also be seen as a parameter change under the same equivalence class of theories. GR is clearly not invariant under the action of this abelian group and therefore cannot be reached by the limit $\omega_{BD}\rightarrow\infty$. Thus, one should not simply assume that BD approaches GR for very large values of the parameter $\omega_{BD}$.

For GBD the situation is more involved since $\omega$ is no longer a parameter but a function of the scalar field. Furthermore, the scalar field dynamic equation \eqref{FulleqKG} has an extra term given by the derivative of the coupling function with respect to the scalar field. This equation can be recast as
\begin{equation}\label{GRlimitKG1}
\nabla_\mu\left(\sqrt{2\omega+3 }\ \nabla^\mu\varphi\right)=\frac{T^{(m)}}{\sqrt{2\omega+3 }}\ .
\end{equation}

The coupling function is arbitrary apart from some boundary conditions. A necessary condition to have a GR-like regime is that when $\varphi$ approaches unit $\omega$ increases boundlessly. Thus, equation \eqref{GRlimitKG1} gives
\begin{equation}\label{GRlimitKG2}
\lim_{\varphi\rightarrow 1} \nabla^\mu\varphi=\frac{\mathcal{A}^\mu}{\sqrt{2\omega+3 }}+\mathcal{B}^\mu +\mathcal{O}\left(\frac{1}{\omega}\right)\ ,
\end{equation}
where $\mathcal{A}^\mu$ and $\mathcal{B}^\mu$ do not depend on the scalar field. For vanishing $\mathcal{B}^\mu$ and constant $\omega$ one recovers equation \eqref{BDeq13}. However, the nontrivial dependence of the coupling function on the scalar field prevents establishing the asymptotic behavior of the scalar field without specifying the coupling function. Notwithstanding, equation \eqref{GRlimitKG2} suffices to show that the last term of equation \eqref{FulleqBD} does not vanish in the corresponding limit. Furthermore, this result does not depend on the nature of the matter content inasmuch as the dominant term on the right-hand side of equation \eqref{FulleqKG} is the derivative of the coupling function with respect to the scalar field.

\begin{figure}
	\includegraphics[width=8cm,height=4.5cm]{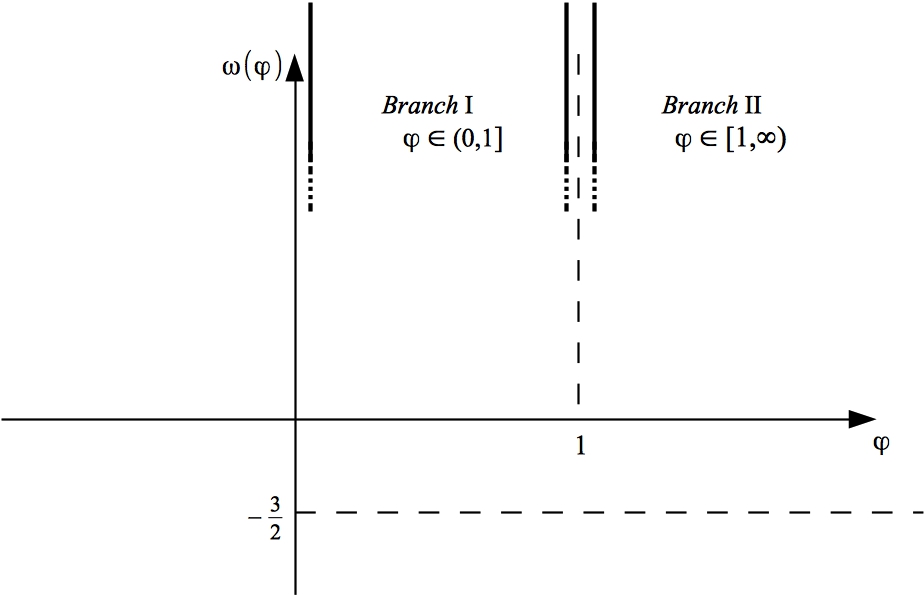}
	\caption{Asymptotic behavior of the coupling function $\omega(\varphi)$ that preserve three physically reasonable conditions: gravitation strength is always non-negative $(\varphi\geq 0)$; the coupling function satisfies $2\omega+3>0$ in order to avoid the scalar field velocity singularity; the system approach asymptotically the General Relativity regime for $\varphi\rightarrow 1$.}\label{figcouplingfunc}
\end{figure}

For a FLRW spacetime, equations \eqref{GRlimitKG1} and \eqref{GRlimitKG2} are ordinary time differential equations. While $\omega\rightarrow\infty$ the scalar field velocity goes to zero, hence we have to analyzed carefully this limit for the product $\omega\,  \dot{\varphi}$. For a finite scale factor, it can be shown that $\omega\,  \dot{\varphi}^2 \propto a^{-6}$ (see eq.~\eqref{KGinteq}) and the dynamic equations reduce to
\begin{align}
	&H^2=\frac{\kappa\rho}{3}+\frac{\kappa\rho_\ast}{3}\left(\frac{a_0}{a}\right)^6-\frac{k}{a^2}\label{FriedGR1}\\
	&\frac{\ddot{a}}{a}=-\frac{\kappa(\rho+3p)}{6} -\frac{2\kappa\rho_\ast}{3} \left(\frac{a_0}{a}\right)^6\label{FriedGR2}
\end{align}
where $\rho_\ast\equiv \frac14 (2\omega_0+3)\dot{\varphi}_0^2$ and $\omega_0$ and $\dot{\varphi}_0$ are constants associated with the initial values of the coupling function and the scalar field velocity. Unless $\dot{\varphi}_0$ is strictly zero, there is always a stiff matter type fluid correction in the GR-like regime. Furthermore, the coupling function has to increase boundlessly when $\varphi$ goes to unit, which can be accounted for only if there is a discontinuity in the coupling function. Thus, there are two disjoint branches as shown in figure \eqref{figcouplingfunc}. For branch I, the GR limit is attained for $\varphi\rightarrow 1^-$ and $\frac{\dd \omega}{\dd \varphi}>0$, while for branch II $\varphi\rightarrow 1^+$ and $\frac{\dd \omega}{\dd \varphi}<0$.

\section{Matching Conditions and Geometrical Setting}\label{MCGS}

The singularity theorems developed by Penrose, Hawking and collaborators \cite{penrose1965,penrose1969,hawking1966,hawking1970,
hawking1965,hawking1973} prove that under physical reasonable conditions, singularities will unavoidably form within GR dynamics. These theorems are general in the sense that they do not suppose any spacetime symmetry. A considerable simpler case is to study the formation of singularities in homogeneous and isotropic solutions. Friedmann-Lema\^itre-Robertson-Walker (FLRW) solutions can be used to describe cosmological and collapsing compact objects solutions. 

Using a spherical coordinate system, the FLRW line element reads
\begin{equation}\label{flrwmetric}
\dd s^2=\dd t^2-a^2(t)\left[\frac{\dd r^2}{1-kr^2}+r^2\left(\dd \theta^2+\sin^2(\theta)\, \dd \phi\right)\right] \ ,
\end{equation}
where $k=0,\pm 1$ defines the spatial section curvature. Conservation of energy-momentum for a barotropic fluid (eq.'s \eqref{consvT} and \eqref{eqstate}) implies
\begin{equation}
\rho=\rho_0\left(\frac{a_0}{a}\right)^{3\gamma}
\end{equation}
with $\rho_0$ and $a_0$ the initial values of matter density and scale factor, respectively. Inserting the FLRW metric in \eqref{FulleqBD}-\eqref{FulleqKG}, the dynamics for a homogeneous and isotropic metric reads
\begin{align}
	&H^2=\frac{\kappa \rho_0}{3\varphi}\left(\frac{a_0}{a}\right)^{3\gamma}+\frac{\omega}{6}\left(\frac{\dot{\varphi}}{\varphi}\right)^2-H\frac{\dot{\varphi}}{\varphi}-\frac{k}{a^2}\label{Friedeq1}\\
	&\frac{\ddot{a}}{a}=-\frac{\left(3+(3\gamma-2)\omega\right)}{2\omega+3}\frac{\kappa \rho_0}{3\varphi}\left(\frac{a_0}{a}\right)^{3\gamma}+H\frac{\dot{\varphi}}{\varphi}\nonumber\\
	&\quad \qquad -\frac{\omega}{3}\left(\frac{\dot{\varphi}}{\varphi}\right)^2+\frac{\dot{\varphi}^2}{2\varphi(2\omega+3)}\frac{\dd \omega}{\dd \varphi}\label{Friedeq2}\\
	&\ddot{\varphi}+3H\dot{\varphi}=\frac{1}{2\omega+3}\left[(4-3\gamma)\kappa \rho_0\left(\frac{a_0}{a}\right)^{3\gamma}-\dot{\varphi}^2\frac{\dd \omega}{\dd \varphi}\right]\label{KGeq}
\end{align}
where we have introduced the Hubble factor $H\equiv \dot{a}a^{-1}$. It is useful to rewrite the Klein-Gordon-like (KGL) equation as 
\begin{equation}\label{KGinteq}
\dot{\varphi}=
\frac{(4-3\gamma)\kappa\rho_0a_0^{3\gamma}}{a^3\sqrt{2\omega+3}}\int \frac{\dd t}{\sqrt{2\omega+3}\, {a}^{3(\gamma-1)}}+
\left(\frac{a_0}{a}\right)^3\sqrt{\frac{2\omega_0+3}{2\omega+3}}\dot{\varphi}_0
\end{equation}
where $\dot{\varphi}_0$ is the initial value of the scalar field velocity. 

Each coupling function $\omega(\varphi)$ defines a different generalized Brans-Dicke theory. This function is arbitrary but must respect a few constraints. Equation  \eqref{KGinteq} shows that $\omega=-3/2$ is a singular point, hence we shall impose $2\omega+3>0$. When $\varphi=1$ the strength of the gravitational field equals its value in GR. Furthermore, the dynamics gets close to GR when the coupling function becomes very large, i.e. $\omega \gg 1$. In addition, the scalar field is associated with the gravitational strength which makes sense only if $\varphi\geq0$. Thus, we shall impose boundary conditions such that (see fig.~\ref{figcouplingfunc})
\begin{align}\label{CFBcond}
\lim_{\varphi\rightarrow 0}{\omega}(\varphi)=\infty \ ,\quad
\lim_{\varphi\rightarrow 1}{\omega}(\varphi)=\infty \ ,\quad \mbox{and}\quad 
2\omega+3>0\ .
\end{align}

A FLRW metric can be used to describe the interior of a collapsing compact object. However, a complete solution requires specifying the spacetime geometry everywhere. Thus, the use of FLRW metric to model collapsing objects needs matching conditions to connect with the exterior solution. Two spacetimes can be smoothly joined by the Israel-Darmois \cite{israel1966,darmois1927} matching conditions. We shall impose that the first and second fundamental forms, namely, the induced metric and the extrinsic curvature are continuous at the boundary hypersurface
\begin{align}\label{mc}
h^{(ext)}_{ab} = h^{(int)}_{ab} \quad , \qquad 
K^{(ext)}_{ab} = K^{(int)}_{ab} \quad . 
\end{align}

For a dust collapsing cloud ($\gamma=1$), the exterior spacetime is described by the Schwarzschild metric~\cite{oppenheimer1939} written in the Eddington-Finkelstein coordinates
\begin{equation}\label{metschw}
ds^{2}_{(ext)} =  \left( 1 - \frac{2M}{\chi} \right) \dd v^{2} - 2 \dd v \dd \chi - \chi^{2} \dd \Omega^{2}
\end{equation}
where $v$ and $\chi$  are, respectively, the advanced and radial coordinates. The interior spacetime is the FLRW metric \eqref{flrwmetric}. Using the matching conditions \eqref{mc}, we arrive at a dust collapsing equation of motion
\begin{equation}
\left(  \frac{\dot{a}}{a} \right)^{2} + \frac{k}{a^2} = \frac{2GM}{a^3R^{3}} \quad , 
\end{equation}
where $M$ is the total mass of the collapsing system. For other fluids, the exterior solution cannot be Schwarzschild since we have a nonzero pressure at the boundary. In these cases, the appropriate exterior spacetime is the generalized Vaidya spacetime \cite{anzhong1999,joshi1999,bojowald2005,bojowald2010}
\begin{equation}\label{vaidyagen}
ds^{2}_{(ext)} =  \left( 1 - \frac{2 M(v,\chi)}{\chi} \right) \dd v^{2} - 2\dd v \dd \chi - \chi^{2} \dd \Omega^{2}\ .
\end{equation}

In the Vaidya line element, the mass function depends only on the advanced coordinates while in the generalized version we have $M(v,\chi)$, hence, the $\partial_\chi M\neq 0$ allows a nonzero pressure at the boundary. We can match all FLRW interior solution with $1 \leq \gamma \leq 4/3$ to the generalized Vaidya exterior solution. The matching can be done by writing the FLRW metric in isotropic coordinates
\begin{equation}\label{metflrwiso}
ds^{2}_{(int)} = \dd t^{2} - \frac{a^{2}(t)}{\left( 1+ {kR^{2}}/{4}  \right)^{2}} \left(\dd R^2+R^2\dd \Omega^{2}\right) \ .
\end{equation}

For these cases, the Israel-Darmois matching conditions read
\begin{align}
\chi =& R a\left(1+\frac{kR^{2}}{4}\right)^{-1}
\ , \label{eqcolrad1} \\
\dot{v} =&\left(1+\frac{kR^{2}}{4} \right) \left(1-\frac{kR^{2}}{4}-R \dot{a}\right)^{-1}\ , \label{eqcolrad2}\\
2M = &aR^{3} \left(\dot{a}^{2}+k\right) \left(1+ \frac{kR^{2}}{4}\right)^{-1}\  , \label{eqcolrad3} \\
\frac{\partial M}{\partial v} =& -\frac{\partial^2 \chi}{\partial v^2} - \left( 1 - \frac{2M}{\chi} - \frac{\partial \chi}{\partial v} \right) \left( \frac{M}{\chi} - \frac{\partial M}{\partial \chi}  \right)\ . \label{eqcolrad4}
\end{align}

For an interior FLRW metric, the exterior spacetime compatible with the junction conditions (\ref{mc}) are the Schwarzschild (\ref{metschw}) or the generalized Vaidya metric (\ref{vaidyagen}). In what follows we shall concentrate on the interior FLRW solution since our concern is the possible formation of spacetime singularities.

\section{Collapsing Compact Objects}\label{CCO}

Our main goal is to study the dynamic system \eqref{Friedeq1}-\eqref{KGeq} and see if a collapsing object can avoid the singularity through the influence of the extra geometrical scalar field. More precisely, we shall impose initial conditions compatible with GR and search for possible dynamics that could avoid gravitational singularities. A direct inspection of \eqref{Friedeq1} shows that a bounce $(H=0)$ can happen only if $\omega<0$. However, GR initial conditions implies $\omega(\varphi_0) \gg 1$ and $|\dot{\varphi}_0|\ll 1$, hence, in order to avoid the singularity, the dynamics has to move far away from GR. Assuming initial conditions compatible with GR for a collapsing compact object means that we have $H<0$ with finite initial size ($a_0$ is finite). Since $\omega \gg 1$, we can neglect the matter contribution to the KGL equation, namely, the scalar field dynamics reads
\begin{equation}\label{CCOBIKG}
\ddot{\varphi}=-3H\dot{\varphi}-\frac{\dot{\varphi}^2}{2\omega+3}\frac{\dd \omega}{\dd \varphi}\ .
\end{equation}
Thus, initially we have
\begin{equation}\label{CCOBIKGint}
\dot{\varphi}=\sqrt{\frac{2\omega_0+3}{2\omega+3}}\left(\frac{a_0}{a}\right)^3 \dot{\varphi}_0\quad .
\end{equation}
Note that for a pure barotropic radiation collapse, equation \eqref{CCOBIKGint} is precise since the matter contribution to the KGL equation vanishes. Furthermore, as long as we can neglect the matter content in the KGL equation, the scalar field velocity does not change sign. If the scalar field starts increasing (decreasing), while the neglected matter term approximation remains valid, it will continue increasing (decreasing) its value.

The evolution of the system depends, evidently, on the coupling function that defines the GBD theory. In order to keep the discussion as general as possible, we shall impose only the boundary conditions \eqref{CFBcond}. Thus, there are two distinct behaviors depending on the range of the scalar field. In branch-I, the scalar field can take values on the interval $\varphi \in \ \left(0,1\right]$. Branch-II covers the complementary domain given by $\varphi \in \ \left[1,\infty\right)$. Let us analyze separately each branch.

\subsection{Branch-I}\label{CCOBI}

\subsubsection*{Case I-a: $\dot{\varphi}_0>0$}\label{CCOBIa}

In this branch, close to $\varphi=1$ we have $\frac{\dd \omega}{\dd \varphi}\gg 1$. Equation \eqref{KGinteq} shows that if $\dot{\varphi}_0>0$ then the scalar field climbs up the coupling function. The dynamics pushes the system even further away from the $\omega<0$ region, hence, there is no bounce. The scale factor contracts continuously until the collapse produces a singularity. Even though the velocity field does not change sign, its acceleration depends on the strength of the velocity field. The manner in which the system approaches the singularity depends on the relative dynamics of the scalar field and a combination of the Hubble factor and the coupling function. Equation \eqref{CCOBIKG} can be written in a more suggestive way as 
\begin{equation}\label{CCOBIKGCII}
\ddot{\varphi}=-3H\dot{\varphi}\left(1-\frac{\dot{\varphi}}{\dot{\varphi}_c}\right)\ ,
\end{equation}
where we have defined a critical velocity 
\begin{equation}
\dot{\varphi}_c\equiv \left|
\frac{1}{3H(2\omega+3)}
\left( \frac{\dd \omega}{\dd \varphi}\right)
\right|^{-1} \ . 
\end{equation}

If the velocity is smaller than the critical value, $\dot{\varphi}<\dot{\varphi}_c$, equation \eqref{CCOBIKGCII} shows that $\ddot{\varphi}>0$, which increases the scalar field velocity. If $\dot{\varphi}>\dot{\varphi}_c$, then $\ddot{\varphi}<0$, which decreases the velocity of the scalar field. As far as we can neglect the matter term, we have three possible situations, depending on the dynamics of $\dot{\varphi_{c}}$ with respect to the velocity of the scalar field. In order to see this behavior, let us approximate the coupling function and its derivative. Close to the divergent point the coupling function can be approximated by
\begin{equation}\label{CCOBIKGCIIIx}
\lim_{\varphi\rightarrow 1^-} \omega=\frac{\omega_\ast}{(1-\varphi)^{n}}\ ,
\end{equation}
where $\omega_\ast>0$ and $n \in \mathbb{N}$. Thus, we have
\begin{align}\label{CCOBIKGCIII}
	\lim_{\varphi\rightarrow 1} \dot{\varphi}&=\sqrt{\frac{\omega_0}{\omega_\ast}}\left(\frac{a_0}{a}\right)^3\dot{\varphi}_0(1-\varphi)^{n/2}\ ,\\
	\lim_{\varphi\rightarrow 1} \frac{\dot{\varphi}}{\dot{\varphi}_c}&=\sqrt{\frac{\omega_0}{\omega_\ast}}\frac{n\dot{\varphi}_0}{6H}\left(\frac{a_0}{a}\right)^3(1-\varphi)^{(n-2)/2}\ . \label{CCOBIKGCIIIa}
\end{align}

\begin{figure}[h]
\includegraphics[width=7cm, height=4.5cm]{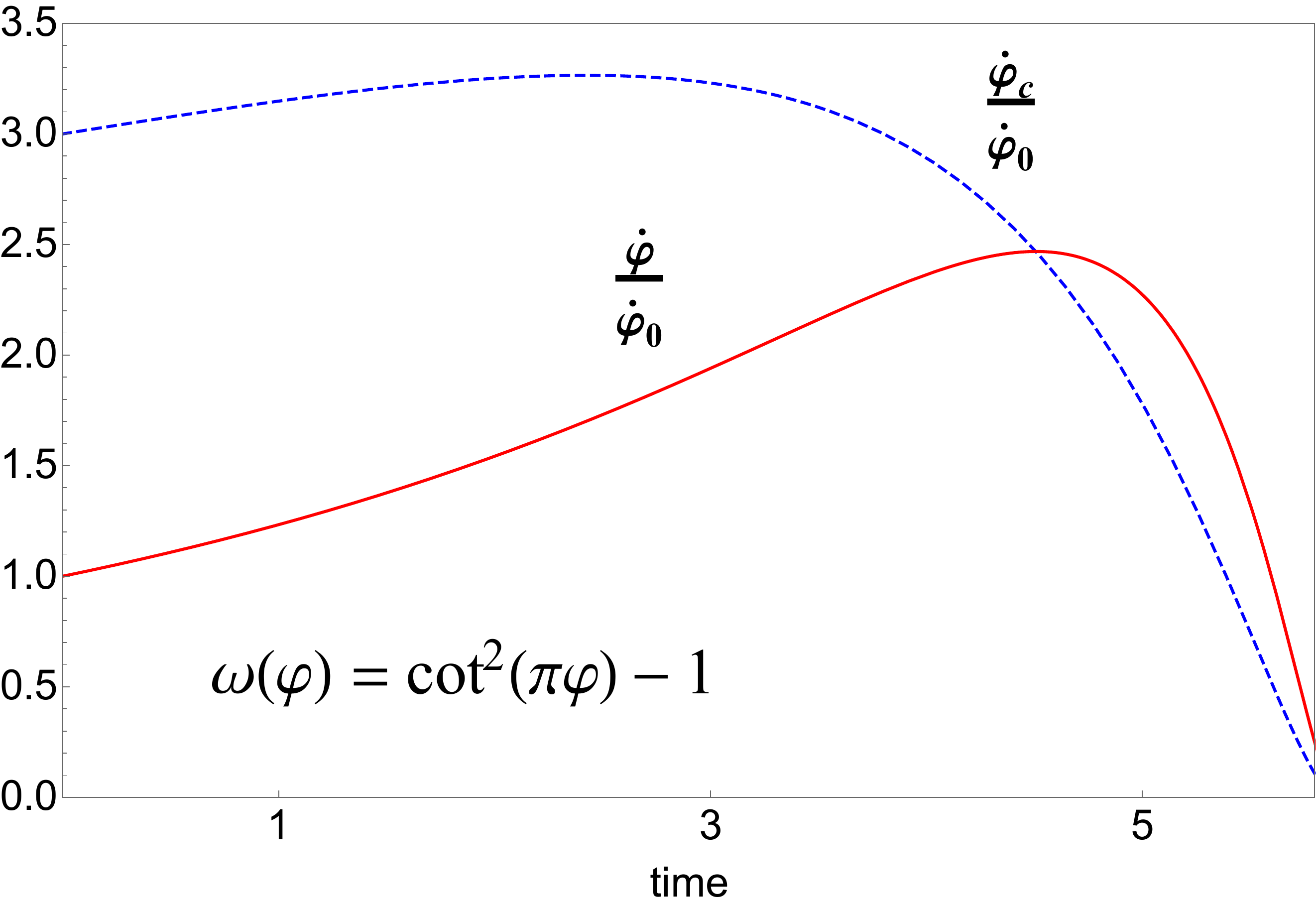}
\caption{Numerical evolution for the initial values satisfying $\dot{\varphi}_{c}(t_0)>\dot{\varphi}_0$. The dashed curve correspond to the evolution of $\dot{\varphi}_{c}$ while the solid curve to $\dot{\varphi}$. As long as $\dot{\varphi}_{c}>\dot{\varphi}$, the scalar field acceleration is positive. But, for $\dot{\varphi}_{c}<\dot{\varphi}$ the acceleration becomes negative and decreases the scalar field velocity.}\label{Ia1phipontocphiponto}
\end{figure}

\begin{figure}[h]
\includegraphics[width=7cm, height=4.5cm]{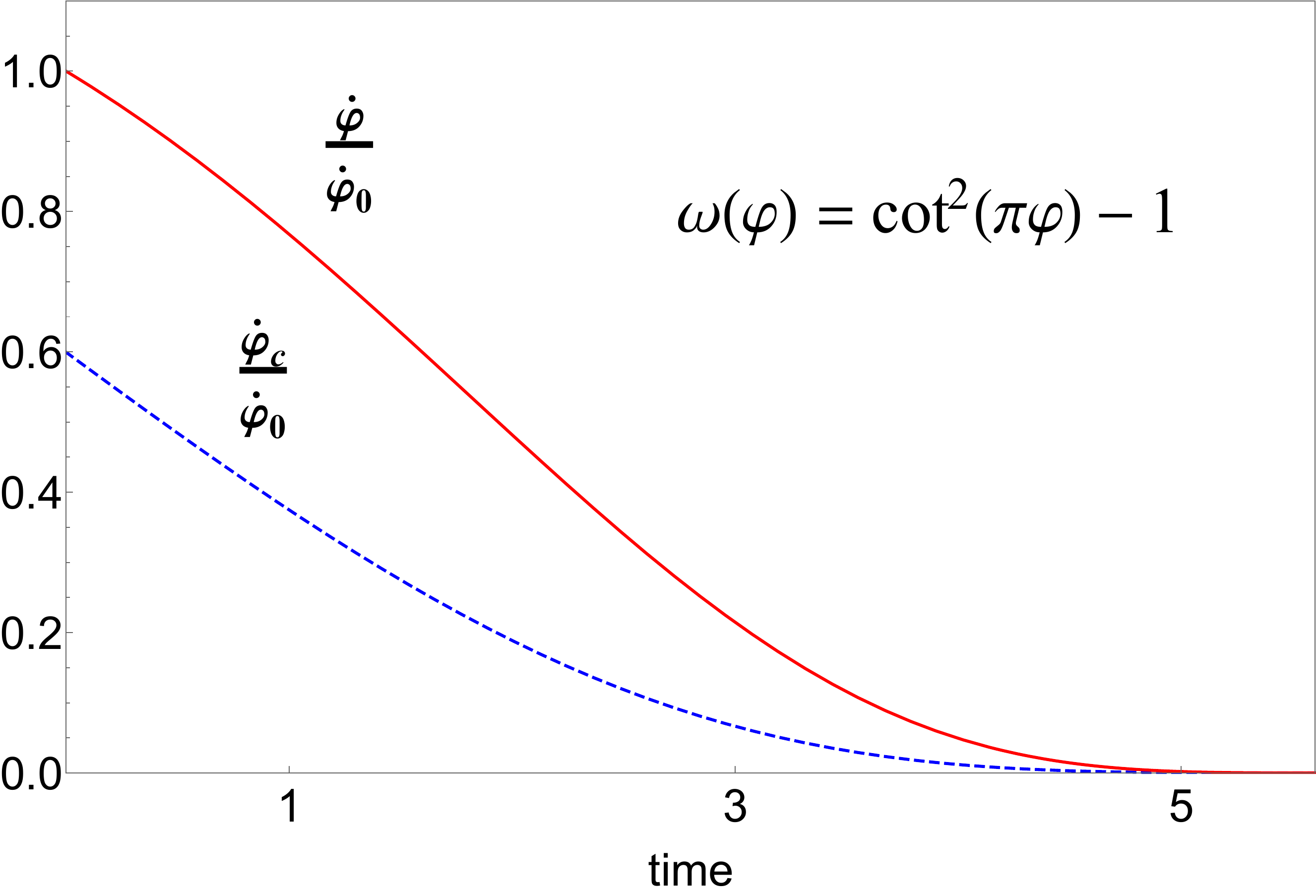}
\caption{Numerical evolution for the initial values satisfying $\dot{\varphi}_0>\dot{\varphi}_{c}(t_0)$. The dashed curve correspond to the evolution of $\dot{\varphi}_{c}$ while the solid curve to $\dot{\varphi}$. The dynamics of $\dot{\varphi}_{c}$ is slow enough that $\dot{\varphi}$ has time to catch-up and the scalar field acceleration goes to zero.}\label{Ia2phipontocphiponto}
\end{figure}

Depending on the power $n$ of the coupling function we can have three different scenarios. When the evolution of $\dot{\varphi}_c$ is slow enough, $\dot{\varphi}$ oscillates around the value of $\dot{\varphi}_{c}$. For a fast varying $\dot{\varphi}_c$, the scalar field velocity $\dot{\varphi}$ never catches up with the evolution of $\dot{\varphi}_{c}$. The third evolution is simply a mixture of both dynamics. Figures (\ref{Ia1phipontocphiponto}) and (\ref{Ia2phipontocphiponto}) display typical behaviors for these scenarios.

\subsubsection*{Case I-b: $\dot{\varphi}_0<0$}\label{CCOBIb}
Let us assume that $\dot{\varphi}_0<0$. In this case, at least initially, we have $\ddot{\varphi}<0$, hence the modulus of the scalar field velocity increases as long as $\omega>0$. The dynamics decreases the value of the scalar field while pushing down the value of the coupling function. 

\begin{figure}[h]
\includegraphics[width=7cm, height=5cm]{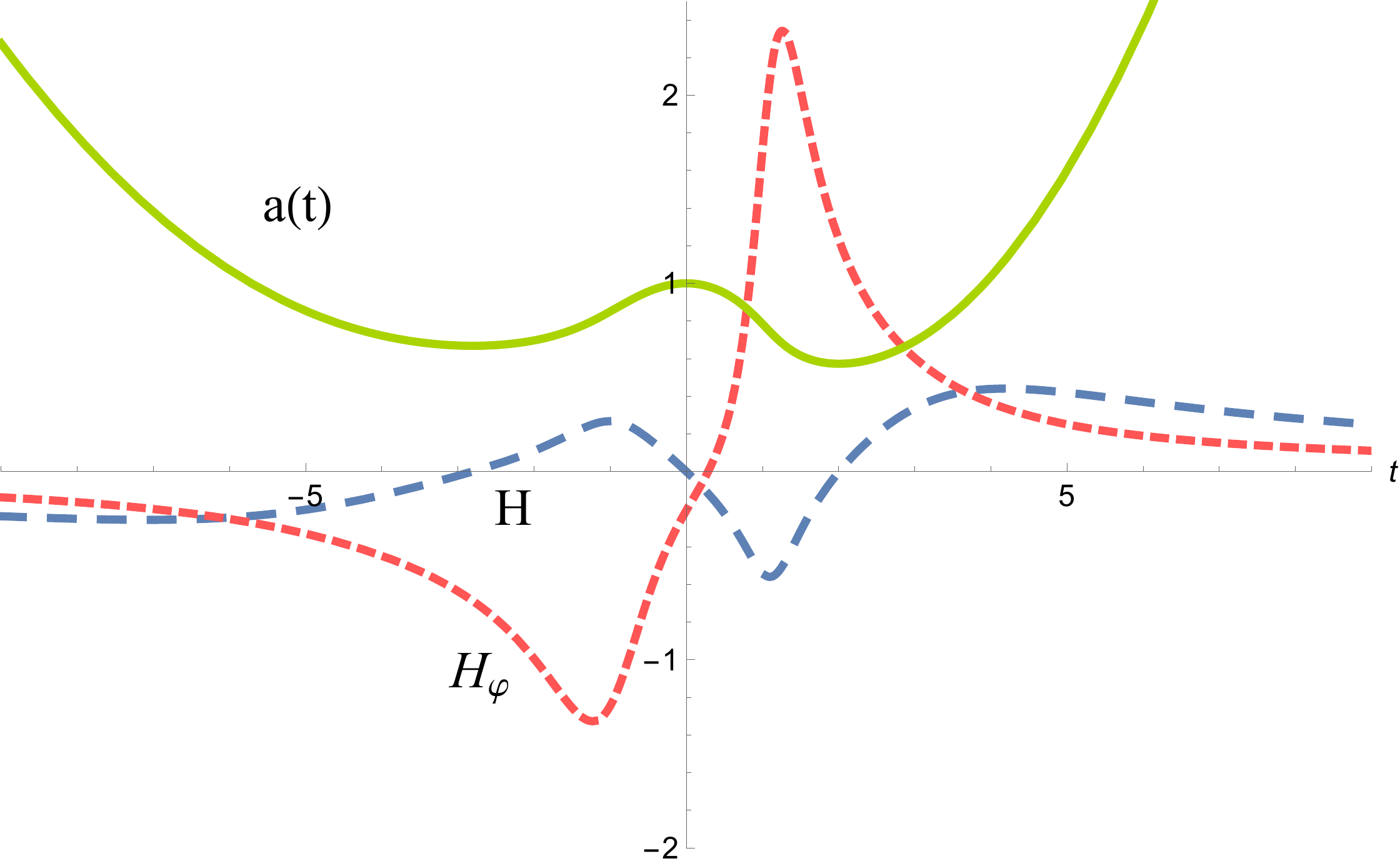}
\caption{Dynamics of the system close to the bounce for a flat negative coupling function ($\omega=-1.4$). The solid curve shows the scale factor bounce while the small-dashed and large-dashed curves show respectively the Hubble parameter for the scale factor and the scalar field $(H_\varphi\equiv {\dot{\varphi}}/{\varphi}$).}\label{DynConstw}
\end{figure}

In this regime, the scalar field contribution to Friedmann's equations is negligible until close to the collapse where \eqref{CCOBIKGint} is no longer valid. In principle, it is possible to choose appropriately the coupling function such that the scalar field reaches the $\omega<0$ region before the collapse. The probability of a bounce increases with the area of the region with negative values of $\omega$. Therefore, let us consider a flat potential well behavior for the coupling function, which means we can neglect the $\frac{\dd \omega}{\dd \varphi}$ term close to the bounce. Numerical evaluation shows that indeed there is a bounce with a negative constant coupling function. Figure \eqref{DynConstw} displays a typical behavior of this kind of solution. Notwithstanding, in order to have a full physical solution we need to connect the bounce behavior of figure \eqref{DynConstw} with the asymptotic regime $\omega \gg 1$.

Let us include Friedmann's equations in our analysis. At the bounce the curvature term is negligible, hence equation \eqref{Friedeq1} gives
\begin{equation}\label{velbBI}
\dot{\varphi}_{bounce}^2=\kappa \rho_0\left(\frac{2\varphi_{bounce}}{|\omega|}\right)\left(\frac{a_0}{a_{bounce}}\right)^{3\gamma}\ .
\end{equation}

To get an order of magnitude of the scalar field velocity at the bounce, we can use the Sun's astrophysical parameters. A compact object initially with the mean density of the Sun has $\kappa \rho_0\approx2,5\times 10^{-6}$. If the length of the system starts roughly with the Sun's volumetric radius and the bounce happens before the Sun's Schwarzschild radius, then $a_0/a_{bounce}\approx2,4 \times 10^5$. Given that $\varphi_{bounce}$ and $|\omega|$ are of the order of unit, and $1\leq \gamma\leq 4/3$, the scalar field velocity is of the order of $\dot{\varphi}_{bounce} \sim 10^5$. 

The range of values of the scalar field in branch I is $\varphi \in \big(0,1\big]$, hence this type of bounce is extremely fine-tuned. The coupling function and the scalar field velocity have to vary orders of magnitude while the scalar field remains practically constant along the collapse. This is possible only if the coupling function is so steep that the deviation from GR happens almost instantaneously. Furthermore, the range of the scalar field shows that everything has to be finely adjusted to hit the $\omega<0$ region with the correct velocity \eqref{velbBI}. A little bit less or more in the initial conditions makes the coupling function never reaches the $\omega<0$ region or leave it too soon as $\varphi\rightarrow0$. Therefore, even though mathematically possible this bounce is physically unreasonable.

\subsection{Branch-II}\label{CCOBII}

There are two main differences from branch-I to branch-II. Initially the derivative of the coupling function is negative, i.e. $\frac{\dd \omega}{\dd \varphi}<0$ and in principle there is no upper limit to the value of the scalar field. Apart from this, we still have $H<0$ with finite initial size ($a_0$ is finite), and since $\omega \gg 1$ we can neglect the matter contribution to the KGL equation
\begin{equation}\label{CCOBIKG2}
\ddot{\varphi}=-3H\dot{\varphi}-\frac{\dot{\varphi}^2}{2\omega+3}\frac{\dd \omega}{\dd \varphi}\ .
\end{equation}

Analogous to branch-I, we can divide branch-II in two cases.

\subsubsection*{Case II-a: $\dot{\varphi}_0<0$}\label{CCOBIIa}

This case is similar to case I-a. As long as we can neglect the matter term we have 
\begin{equation}\label{CCOBIIKGCI}
\ddot{\varphi}=-3H\dot{\varphi}\left(1+\frac{\dot{\varphi}}{\dot{\varphi}_c}\right)\ ,
\end{equation}
where the critical velocity is defined strictly positive as before and the change of sign is due to the change of sign of the derivative of the coupling function. In a collapsing system, for a negative initial velocity we have $H\dot{\varphi}_0>0$, hence if $\left|\dot{\varphi}\right|<\dot{\varphi}_c$ the acceleration is negative, which increases the modulus of the scalar field velocity. On the other hand, if $\left|\dot{\varphi}\right|>\dot{\varphi}_c$, then the acceleration is positive. Once more, while the approximation we have three possible behaviors as depicted on case I-a. 

There is, however, one important difference when the matter term starts to dominate. Contrary to the case I-a, the positive acceleration produced by the matter term reverse the scalar field dynamics pushing it down to the $\omega<0$ region. The crucial point is now to analyze if the system can reach the bounce allowed region before the scale factor goes to zero. Once again we need to take into account Friedmann's equations.

Close to the singularity, when the matter term dominates, Friedmann's equation \eqref{Friedeq1} can be approximated to 
\begin{equation}\label{CCOBIIFried1CI}
H^2\approx\frac{\kappa \rho_0}{3\varphi}\left(\frac{a_0}{a}\right)^{3\gamma}+\frac{\omega}{6}\left(\frac{\dot{\varphi}}{\varphi}\right)^2\ ,
\end{equation}
and for positive coupling function we have $\|H\| \geq \|H_\varphi\|$. Thus the velocity of the scale factor is greater than the scalar field's. A comparison of the second Friedmann's equation \eqref{Friedeq2} with KGL \eqref{KGeq} also shows that the scale factor acceleration is greater than the scalar field's. In the limit $a\rightarrow 0$ we can approximate
\begin{align}
	&\frac{\ddot{a}}{a}\approx-\frac{\left(3+(3\gamma-2)\omega\right)}{2\omega+3}\frac{\kappa \rho_0}{3\varphi}\left(\frac{a_0}{a}\right)^{3\gamma}-\frac{\omega}{3}\left(\frac{\dot{\varphi}}{\varphi}\right)^2\label{CCOBIIFried2CI}\\
	&\frac{\ddot{\varphi}}{\varphi}\approx\frac{(4-3\gamma)}{2\omega+3}\kappa \rho_0\left(\frac{a_0}{a}\right)^{3\gamma}\label{CCOBIIKGCI2}
\end{align}

For positive value of the coupling function, the above equations show that $\left|\frac{\ddot{a}}{a}\right|\geq \left|\frac{\ddot{\varphi}}{\varphi}\right|$. Even though the system does move in the direction of the bounce allowed region, the dynamics hits the singularity $a=0$ before the scalar field reaches the $\omega<0$ region. Thus, there is also no bounce for this case and the singularity is unavoidable.

\subsubsection*{Case II-b: $\dot{\varphi}_0>0$}\label{CCOBIIb}

This is the most promising scenario for a bounce since the system can be driven to the bounce allowed region right from the beginning. Since $d \omega / d \varphi <0$ and $\dot{\varphi}_0>0$, the right-hand side of equation \eqref{KGeq} is positive, hence, during the collapse the sign of $\ddot{\varphi}$ is always positive. Similarly to case II-a, Friedmann's equations show that the scale factor dynamics is faster than the scalar field's. During the contracting phase, we can recast equations \eqref{Friedeq1}-\eqref{Friedeq2} as two inequalities
\begin{align}
H^2-\frac{\omega}{6}H_\varphi^2&>0\ ,\label{CCOBIIFried1CIIb}\\
\frac{\ddot{a}}{a}+\frac{\ddot{\varphi}}{3\varphi}&<0\ . \label{CCOBIIFried2CIIb}
\end{align}

These relations show that indeed the scale factor velocity and acceleration are greater than the scalar field velocity and acceleration. Again, the issue is if the system has enough time to reach the $\omega<0$ region. The coupling function is arbitrary, which allows for fast variations of $\omega$ even with small variation of $\varphi$ as long as $\left|\frac{\dd \omega}{\dd \varphi}\right|>>1$. Inequalities \eqref{CCOBIIFried1CIIb}-\eqref{CCOBIIFried2CIIb} give the relative dynamics of $a$ and $\varphi$ but we need to compare how much $\omega$ varies with the scale factor. As long as we can arbitrary choose the coupling function we can always find a function that allows for the system to reach the $\omega<0$ region before the singular point at $a=0$. Notwithstanding a closer analysis shows that even in this case there is no bounce.

At the bounce (if possible), combining equations \eqref{Friedeq1}-\eqref{Friedeq2} one can show that
\begin{align}\label{CCOBIIFried2CIIb2}
\frac{\ddot{a}}{a}=\frac{3\left[1-(\gamma-2)\omega\right]}{2\omega+3}\frac{\kappa \rho_0}{3\varphi}\left(\frac{a_0}{a}\right)^{3\gamma}\qquad .
\end{align}

A bounce means $\ddot{a}>0$, which using equation~\eqref{CCOBIIFried2CIIb2} translates into the condition $0>\omega>-1/(2-\gamma)$. In order to have a bounce, the coupling function has only a narrow window of at most $\omega \in \left(0,-1\right)$. As a consequence, we also need a small derivative of the coupling function. Otherwise the system would leave this region too soon. This condition also has the advantage to increase the probability of a bounce since, once in the $\omega<0$ region, the system will remain there until a possible bounce occurs.

KGL equation \eqref{KGeq} guarantees that the velocity field only increases during the collapse and is always positive. This feature, that at first sight could favor a bounce, actually invalidates it. We can rewrite Friedmann's equation \eqref{Friedeq1} as
\begin{align}\label{CCOBIIFried1CIIb2}
\frac{\kappa \rho_0}{3}\left(\frac{a_0}{a}\right)^{3\gamma} - \frac{k\varphi}{a^2} = \dot{\varphi}\left[ -\frac{\omega}{6}\frac{\dot{\varphi}}{\varphi}+H\right]+H^2\varphi 
\end{align}

At the bounce the scale factor reaches its minimum, hence the left-hand side of equation \eqref{CCOBIIFried1CIIb2} reaches its maximum. Since the second time derivative of the scalar field has to be positive, if $\dot{\varphi}>0$, then the velocity of the scalar field will continue to grow during the bounce. Thus, instead of decreasing, the right-hand side (RHS) of \eqref{CCOBIIFried1CIIb2} increases, which contradicts the hypothesis of a bounce. A bounce with almost constant coupling function can happens only if the velocity of the scalar field is negative, which would allow the RHS of \eqref{CCOBIIFried1CIIb2} to decrease in time.

\subsection*{Summary for Collapsing Compact Object}

We considered all possible cases for a spherically symmetric and homogeneous collapsing object with arbitrary coupling function. We divided our analysis in four cases. In both cases I-a and II-a the dynamics has $\omega \rightarrow \infty$ as an attractor. In case I-a, this feature remains during the whole evolution while in case II-a is an attractor only till the matter term starts to dominate in the KGL equation. Thereafter, the system moves away from this regime but has not enough time to reach the bounce allowed region ($\omega<0$) before the singularity at $a=0$. In case I-b the system also moves away from $\omega\gg 1$ right from the beginning and can eventually access the $\omega<0$ region. However, branch-I has a limited range for the scalar field, which makes the bounce fine-tuned. Even though mathematically possible, a bounce in case I-b suffers from fine-tuning. Finally, case II-b can also access the $\omega<0$ region, but a closer analysis of the dynamic shows that a bounce can happen only if the scalar field velocity is negative, which in this case is prohibited by the KGL equation.

\begin{table}[h]
	\begin{center}
		\begin{tabular}{ |c|c|c|c| }
			\hline
			\multicolumn{4}{ |c| }{Summary for Collapsing Compact Objects} \\
			\hline
			 \hspace{0.1cm} Branch \hspace{0.1cm} & \hspace{0.1cm} $\dot{\varphi}$ changes sign \hspace{0.1cm}&\hspace{0.1cm} Access $\omega<0$ \hspace{0.1cm}&\hspace{0.15cm} Singularity \hspace{0.15cm} \\ \hline
			 I-a & no & no & unavoidable \\
			 I-b & yes & yes & fine-tuned \\ \hline
			 II-a & yes & no & unavoidable \\
			 II-b & no & yes & unavoidable \\ 
			\hline
		\end{tabular}
	\end{center}\label{table1}
\end{table}

\section{Cosmology}\label{COSMO}

For the purpose of our study, we shall consider only homogeneous and isotropic cosmological models. They have strong observational support and suffice for our analysis. A FLRW universe can have a collapsing phase and/or an expanding phase. A collapsing FLRW universe is formally identical to the cases examined in the collapsing compact object. Therefore, it suffice to study expanding universes with $H>0$.  We consider only initial conditions close to GR and investigate each branch separately (see Figure~\eqref{figcouplingfunc}).

In an expanding universe the singularities are of the Big-Rip type (BR) \cite{nojiri2005,chimento2004,caldwell2003}. These singularities are characterized by the divergence of spacetime parameters in a future finite time $t_{BR}$, i.e.
\begin{align}
\lim_{t\rightarrow t_{BR}}a(t)=\infty \ , \ 
\lim_{t\rightarrow t_{BR}}H(t)=\infty \ ,  \
\lim_{t\rightarrow t_{BR}} \ddot{a}(t)=\infty \ .
\end{align}

The first Friedmann equation shows that a BR can happen only if $|H_\varphi |$ also goes to infinity. Indeed, it is the scalar field that must drive the BR, and, close to this point, we can approximate equation~\eqref{Friedeq1} to
\begin{equation}\label{FriedeqBR1}
H^2+HH_\varphi\approx \frac{\omega}{6}H_\varphi^2\ .
\end{equation}

The above equation shows that, contrary to the collapsing case, a necessary condition for a BR is $\omega>0$. Let us analyze individually each branch of the coupling function.

\subsection{Branch-I}\label{COSMOBI}

\subsubsection*{Case I-c: $\dot{\varphi}_0>0$}\label{COSMOIa}

There is no novelty in this case. The integrated KGL equation~\eqref{KGinteq} shows that if $\dot{\varphi}_0>0$ then $\dot{\varphi}$ cannot change sign and the system will climb up the coupling function indefinitely. As long as the coupling function increases boundlessly, the contribution of the scalar field can be at most as stiff matter, which decreases with $a^{-6}$, hence GR can be seen as an asymptotic regime.

\subsubsection*{Case I-d: $\dot{\varphi}_0<0$}\label{COSMOIb}

In branch-I, the scalar field is limited to $\varphi \in \left(0,1 \right]$. An initial negative velocity pushes the system down in the coupling function and to small values of the scalar field. There are two ways to produce a BR: the velocity of the scalar field can diverge to plus or minus infinity. Let us first consider the $H_\varphi\rightarrow - \infty$ possibility.

Assuming initial conditions close to GR, the initial velocity of the scalar field has to be small and the coupling function (and its derivative) very large. In order to have a $H_\varphi\rightarrow - \infty$ in finite time, the acceleration of the scalar field has also to diverge to minus infinity. This implies that the matter term in the KGL equation can never dominate and the coupling function derivative has to be positive. Thus, the BR has to occur before the scalar field reaches the $\omega<0$ region. Evidently, this is a very fine-tuned condition and physically unrealizable.

Naively, one might expect that $\varphi\rightarrow 0$ could engender a BR since the right-hand side of the Friedmann equation~\eqref{Friedeq1} is inversely proportional to the scalar field. However, on the left side of branch-I, the scalar field acceleration becomes positive and slow-down the scalar field. It is impossible to produce a BR in the limit $\varphi\rightarrow 0$ if the coupling function diverges ($\omega\gg1$) at the origin.

The second possibility is the scalar field velocity diverges to plus infinity. For this to happen, the scalar field velocity must change sign. At $H_\varphi=0$, the scalar field acceleration is positive ($\ddot{\varphi}>0$) showing that it is a local minimum. After this point, since we are considering an expanding universe with $H>0$, the scalar field will return and climb up the right side of the coupling function and everything goes as in the previous case I-c where there is no BR.

\subsection{Branch-II}\label{COSMOBI}

\subsubsection*{Case II-c: $\dot{\varphi}_0<0$}\label{COSMOIIa}

A negative $\dot{\varphi}_0$ in branch-II means that initially the scalar field moves upward in the coupling function. Let us recall the integrated version of the KGL equation
\begin{equation}\label{KGinteq2}
\dot{\varphi}=
\frac{ (4-3\gamma)\kappa\rho_0 a_0^{3\gamma}}{a^3\sqrt{2\omega+3}}\int \frac{\dd t}{\sqrt{2\omega+3}\, {a}^{3(\gamma-1)}}+
\left(\frac{a_0}{a}\right)^3\sqrt{\frac{2\omega_0+3}{2\omega+3}}\dot{\varphi}_0
\end{equation}

The matter term on the right-hand side above is always positive and, if $\dot{\varphi}_0<0$, the last term is always negative. Initially the matter term is negligible compared with the other term since it has a higher power of the coupling function in the denominator. In addition, the dynamics decreases the matter term even further since $H>0$ and the scalar field is moving upward in the coupling function. Thus, the negative term dominates and the scalar field velocity never changes sign. Once again, the contribution of the scalar field is at most of a stiff matter, which in an expanding universe is negligible if compared with the conventional matter term. Therefore, the system asymptotically approaches a GR regime.

\subsubsection*{Case II-d: $\dot{\varphi}_0>0$}\label{COSMOIIb}

The KGL equation guarantees that an initially positive $\dot{\varphi}_0$ is a sufficient condition for the scalar field to be always positive. Thus, the system moves down the coupling function. A necessary condition for a BR is that $H_\varphi\rightarrow\infty$, see equation~\eqref{FriedeqBR1}. However, the KGL equation shows that for a $\omega>0$ and $a\rightarrow\infty$, we have $H_\varphi\rightarrow 0 $, which contradicts the assumption of a BR. Thus, even though we have a complete freedom in choosing the coupling function, there is also no BR for this case.

\subsection*{Summary for Cosmological Dynamics}

We considered all possible cases for a FLRW universe with an arbitrary coupling function. As before, we divided our analysis in four cases that are summarized in table below. None of the four cases have BR type singularities. Case I-d has a very fine-tuned possibility that it is physically discarded. In terms of dynamics, both cases I-c and II-c have GR as an asymptotic future attractor. 

\begin{table}[h!]
	\begin{center}
		\begin{tabular}{ |c|c|c|c| }
			\hline
			\multicolumn{4}{ |c| }{Summary for Cosmology} \\
			\hline
			\hspace{0.08cm} Branch \hspace{0.1cm} & \hspace{0.08cm} $\dot{\varphi}$ changes sign \hspace{0.1cm}&\hspace{0.08cm} Asymptotic GR \hspace{0.1cm}&\hspace{0.08cm} BR Singularity \hspace{0.1cm} \\ \hline
			I-c & no & yes & no \\
			I-d & possible & possible &  fine-tuned\\ \hline
			II-c & no & yes & no \\
			II-d & no & no & no \\ 
			\hline
		\end{tabular}
	\end{center}
\end{table}


\section{Conclusions}\label{Conclusions}

Generalized Brans-Dicke theories are natural extensions to Brans-Dicke original proposal that maintains the same non-minimal coupling between the curvature and the scalar field while introducing a new coupling function to its kinetic term. The main motivation is to allow the Brans-Dicke parameter to vary in different gravity scenarios, such as, between primordial universe and solar system dynamics. In the present work we have studied the formation of classical singularities in GBD. Given the arbitrariness of the coupling function, one could argue that with an adequate choice of $\omega(\varphi)$, in principle, it would be possible to dynamically avoid classical singularities. The simplest scenarios are homogeneous and isotropic spatial section described by the family of FLRW metrics.

Our analysis depends only on two physically motivated hypotheses. Solar system experiments and cosmological observations show that GR is in good agreement with experimental data\cite{will1993,ade2015.1,ade2015.2}. Thus, we assume initial conditions that mimic a GR regime, namely, small scalar field velocities $\left(|H_{\varphi}|\ll 1\right)$ and large coupling function ($\omega\gg1$). We impose boundary conditions for the coupling function such that the effective gravitation strength, $G_{eff} \sim \varphi$, is non-negative and satisfies the conditions given by figure~\eqref{figcouplingfunc}.

We have shown that spherically symmetric and homogeneous collapsing objects generically form singularities. Case I-b can avoid the singularity but the solution is fine-tuned and not physically realizable. Indeed, it seems reasonable that less symmetric solutions might relax this fine-tuned condition for singularity avoidance; we shall analyze carefully this possibility in a future work. Contracting FLRW cosmological solutions are formally identical to the collapsing cases. Thus, the cosmological singularities that plague GR are also present in GBD. On the other hand, expanding cosmological scenarios are completely free of Big Rip-type singularities. We have shown that in an expanding universe the scalar field behaves at most as a stiff matter type fluid, which makes the system asymptotically approach GR.


\begin{acknowledgments}
	The authors are grateful to Nelson Pinto Neto and Santiago Perez Bergliaffa for useful comments. We would like to thank CNPq of Brazil for financial support.
\end{acknowledgments}



\end{document}